\documentclass[12pt]{article}
\pdfoutput=1
\usepackage{amsmath,amssymb,amsthm,bm,graphicx,float,array,multirow,multicol,rotfloat,caption,subcaption,hyperref,cleveref,enumerate,geometry,mathdots,adjustbox,booktabs,parskip,mathtools,tikz,tikz-cd,pdflscape,csquotes,lscape,rotating,empheq,mathrsfs,cases,xcolor,bm,dsfont}
\usepackage[para]{threeparttable}
\usepackage[all]{xy}
\usepackage[normalem]{ulem}
\usepackage[numbers,sort&compress]{natbib}
\usepackage[shortlabels]{enumitem}
\usepackage[para]{threeparttable}
\usepackage{makecell}
\usetikzlibrary{positioning}
\numberwithin{equation}{section}
\numberwithin{figure}{section}
\allowdisplaybreaks
\makeatletter

\hypersetup{colorlinks=true}
\hypersetup{linkcolor=black}
\hypersetup{citecolor=black}
\hypersetup{urlcolor=black}

\theoremstyle{plain}
\newtheorem*{thm*}{Theorem}

\theoremstyle{definition}

\newtheorem*{defn*}{Definition}

\makeatother

\DeclareMathOperator{\Tr}{Tr}

\newcommand\scalemath[2]{\scalebox{#1}{\mbox{\ensuremath{\displaystyle #2}}}}

\newcommand{\wtr}{\widetilde{r}}
\newcommand{\wtrp}{\widetilde{r}\hspace{.5pt}{}^{\prime}}
\newcommand{\wtrpp}{\widetilde{r}\hspace{.5pt}{}^{\prime\prime}}
\newcommand{\wtrpS}{\widetilde{r}\hspace{.5pt}{}^{\prime}}

\newcommand{\wtrpSsq}{\widetilde{r}\hspace{.5pt}{}^{\prime\,2}}
\newcommand{\wtrppSsq}{\widetilde{r}\hspace{.5pt}{}^{\prime\prime\,2}}
\newcommand{\wtm}{\widetilde{m}}
\newcommand{\wtw}{\widetilde{w}}
\newcommand{\wth}{\widetilde{h}}
\newcommand{\wts}{\widetilde{\sigma}}

\newcommand{\dwtm}{\dot{\widetilde{m}}}
\newcommand{\dwth}{\dot{\widetilde{h}}}
\newcommand{\dwts}{\dot{\widetilde{\sigma}}}

\begin{document}

\begin{titlepage}
\vspace*{-3cm} 
\begin{flushright}
{\tt CALT-TH-2022-020,  IPMU 22-0026}\\
\end{flushright}
\begin{center}
\vspace{2.2cm}
{\LARGE\bfseries A universal formula for the density of states with continuous symmetry \\}

\vspace{1.8cm}

{\large
Monica Jinwoo Kang,$^1$ Jaeha Lee,$^1$ and Hirosi Ooguri$^{1,2}$\\}
\vspace{.8cm}
{$^1$  Walter Burke Institute for Theoretical Physics \\
California Institute of Technology, Pasadena, CA 91125, U.S.A.}
\vspace{.2cm}
\\
{$^2$ Kavli Institute for the Physics and Mathematics of the Universe (WPI) \\
University of Tokyo, Kashiwa 277-8583, Japan}
\vspace{.2cm}

\vspace{1.2cm}
\vspace{.2cm}
\vspace{.2cm}
\textbf{Abstract}
\end{center}
We consider a $d$-dimensional unitary 
conformal field theory with a compact Lie group global symmetry $G$ and show that, at high temperature $T$ and on a compact Cauchy surface, the probability of a randomly chosen state being in an irreducible unitary representation $R$ of $G$ is proportional to $(\operatorname{dim}  R)^2 \ \exp[-  c_2(R) /(b \, T^{d-1}) ]$.
We use the spurion analysis to derive this formula and relate the constant $b$ to a domain wall tension. 
We also verify it for free field theories and holographic conformal field theories and compute $b$ in these cases.
This generalizes the result in \href{https://arxiv.org/abs/2109.03838}{\tt 2109.03838} that the probability is proportional to $(\operatorname{dim} R)^2$ when $G$ is a finite group.
As a by-product of this analysis, we clarify thermodynamical properties of black holes with non-abelian hair in anti-de Sitter space.

\vspace{1cm}
\vfill 
\end{titlepage}

\tableofcontents

\newpage

\section{Introduction}

In \cite{Harlow:2021trr}, a simple formula is derived for the density of black hole microstates 
in theory with finite group gauge symmetry $G$. The formula states that, if we pick a random state from a uniform distribution of all states of the black hole in the semiclassical regime, the probability of it being in a unitary irreducible representation $R$ of $G$ is
\begin{equation}
    P_R = \frac{ ({\rm dim}  R)^2 }{|G|} \,,
    \label{finite}
\end{equation}
where $|G|$ is the number of elements in $G$ so that,
\begin{equation}
    \sum_R P_R = 1 \,.
\end{equation}
It was also conjectured in the paper that the formula applies to any
conformal field theory (CFT) at high temperature on a sphere with finite group global symmetry $G$. This generalizes the result of \cite{Pal:2020wwd} from two dimensions to arbitrary dimensions. The conjecture is verified in the context of free field theories and weakly coupled theories in \cite{Cao:2021euf}, and a general derivation is presented in \cite{Magan:2021myk} using the result of \cite{Casini:2020rgj}.  See also \cite{Coleman:1991ku} for earlier results on black holes with discrete gauge charges in specific models. 

In this paper, we generalize this result to the case where $G$ is a compact Lie group. Since $|G|$ is infinite and $G$ has infinitely many unitary irreducible representations, equation \eqref{finite} needs modifications. We show that, at high temperature and on a compact Cauchy surface, the probability $P_R$ for a random state to be in a representation $R$ of $G$ is given by
\begin{equation}
    P_R=(\operatorname{dim}R)^2\ \left(\frac{4\pi}{b T^{d-1}} \right)^{\dim G/2} \exp\left[ -\frac{c_2(R)}{b\,T^{d-1}} +\cdots \right]  \,
    \label{eqn:holographic} \,,
\end{equation}
where $T$ is the temperature, $d$ is the dimensions of the spacetime of the CFT, 
$c_2(R)$ is the second Casimir of $R$, and $\cdots$ represents terms subleading in $1/T$.
An important point is that $b$ is a positive constant independent of $R$ and $T$.
For small representations, where $c_2(R)\ll T^{d-1}$, the $R$-dependence of $P_R$ is captured by the $(\operatorname{dim}R)^2$ factor as in the finite group case (\ref{finite}). For large representations where $c_2(R)\gg T^{d-1}$, $P_R$ decays exponentially.

We derive equation (\ref{eqn:holographic}) by calculating the twisted partition function,
\begin{equation}
    Z(T,g) = \Tr\left[ U(g)\ e^{-\beta \widehat{H}} \right] \,,
    \label{twistedbyg}
\end{equation}
where the trace is taken over the CFT Hilbert space, $U(g)$ is the action of $g \in G$ on the Hilbert space, $\beta = 1/T$, and $\widehat{H}$ is the Hamiltonian. When $g=1$, it is the standard partition function with the universal large $T$
behavior,
\begin{equation}
    Z(T,g=1) = \exp\left( a\,T^{d-1}+\cdots \right) \,,
\label{eqn:addef}
\end{equation}
for some constant $a$. In two dimensions, it is related to the Cardy formula with
\begin{equation}
    a= \pi^2 (c_L+c_R)/6 \, ,
\end{equation}
where $c_L$ and $c_R$ are the central charges in the left and right movers. 

We employ the spurion analysis for the theory obtained by dimensional reduction of the CFT
on the thermal circle
and show that the $g$ dependence of $Z(T,g)$ is of the form
\begin{equation}
     Z(T,g= e^{i\phi}) = \exp\left( a\,T^{d-1}- \frac{b}{4} T^{d-1} \langle \phi, 
     \phi\rangle +\cdots \right) \,,
    \label{twistedexpansion}
\end{equation}
where the inner product $\langle \phi, 
     \phi\rangle$ is given by the trace of $\phi^2$ in the adjoint representation. 
The constant $b$ is related 
to the tension of the domain wall which
generates the $g$-twisted sector and therefore is positive.
 We also verify this formula by calculating $b$ for free field theories and 
for holographic conformal field theories.
Since the twisted partition function $Z(T,g)$ is a class function of $g$, \emph{i.e.}, invariant under the conjugation $g\rightarrow hgh^{-1}$ for any $h\in G$, we can expand $Z(T,g)$ in characters $\chi_R(g)$
of unitary irreducible representations of $G$. We calculate the coefficients for the expansion of equation (\ref{twistedexpansion}) and obtain
\begin{equation}
    Z(T,g)/Z(T,1) = 
\left(\frac{4\pi}{b T^{d-1}} \right)^{\dim G/2}   
\sum_{R}\ \operatorname{dim}\,R \cdot  \chi_R(g)\ \exp\left(  -\-\frac{c_2(R)}{b\,T^{d-1}}+\cdots \right) \,
   \,.
    \label{CharacterExpansion}
\end{equation}
Our main result (\ref{eqn:holographic}) then follows. 

For $d=1$, equation \eqref{eqn:holographic} is derived
for BF gauge theory coupled to Jackiw--Teitelboim gravity \cite{Kapec:2019ecr}. 
For $d=2$, the formula for $G=U(1)$ is derived using the modular invariance of 2d CFTs \cite{Pal:2020wwd}. 
Our results generalize this to $d \geq 3$ and to non-abelian $G$. 
The exponential suppression factor in equation (\ref{twistedexpansion}) is also mentioned for free field theories in a note added to \cite{Cao:2021euf}.
We note that the right-hand side of equation (\ref{CharacterExpansion}) is in the same form as that of the partition function of the two-dimensional Yang--Mills theory with gauge group $G$ and the coupling constant proportional to $1/T^{(d-1)/2}$ 
\cite{Migdal:1975zg,Rusakov:1990rs,Fine:1990zz,blau:1991mp,Witten:1992xu}. There may also be a connection between our results and the recent study of the entanglement entropy in the presence of a global symmetry \cite{Casini:2019kex}. 

In the holographic derivation of equation \eqref{CharacterExpansion}, we use the Einstein gravity coupled to the Yang--Mills theory with gauge group $G$ and a finite number of matter fields in anti-de Sitter space (AdS). When $G$ is non-abelian, there are two types of relevant bulk geometries besides the thermal AdS: black holes with and without non-abelian hair. Both bulk geometries obey the same boundary condition at the infinity of AdS. However, the former has genuinely non-abelian configurations of the gauge field, while the gauge field in the latter is commutative. There is an extensive literature on such solutions (see \cite{Volkov_1999,Winstanley:2008ac} for some reviews).
One of the outstanding questions in this area has been whether solutions with non-abelian hair are thermodynamically stable. As we will show in this paper, the two types of solutions, with and without non-abelian hair, converge in the high temperature limit $T \rightarrow \infty$. We compute the $1/T$ corrections to their thermodynamical quantities for
purely electric solutions and show that the black holes with non-abelian hair have lower free energies. This determines that the black holes with non-abelian hair are thermodynamically more stable.

The coefficients $a$ and $b$ computed 
for free field theories and holographic CFTs are summarized in Table \ref{tab:coefficients} below. When we have $N$ free scalars or $N$ free fermions, both $a$ and $b$ are proportional to $N$. In holographic CFTs, both $a$ and $b$ are proportional to $\ell^{d-1}/G_N$ assuming $G_N \sim e^2$, where $G_N$ is the Newton's constant, $e$ is the gauge coupling constant, and $\ell$ is the curvature radius of AdS. Thus, in both the free field theories and holographic CFTs, $a$ and $b$ are proportional to the number of degrees of freedom of the system. 

\begin{table}[H]
\begin{threeparttable}
    \centering
    \[\arraycolsep=12pt
    \begin{array}{@{\ }c c c@{\ }}
    \toprule
        & a & b \\\midrule 
        \text{A free scalar with } G=U(1) & 2\zeta(d) &  4\zeta(d-2) \\\midrule
        \begin{array}{@{}c@{}}
        \text{A free scalar in}  \\
        \text{a representation } \rho \text{ of } G
        \end{array} & 2\zeta(d)\dim \rho &
        4\zeta(d-2) \ c_2(\rho)  \dfrac{\dim \rho }{\dim G} \\[10pt] \midrule
        \text{A free spinor with } G=U(1)  & \begin{array}{l c@{}}
            d=2 :\;\; \zeta(2) = \pi^2/6\\[6pt]
            d=3 :\;\qquad 3\zeta(3)
        \end{array} & \begin{array}{c}
            \quad 1 \\[6pt]
            16\log 2
        \end{array} \\[-16pt] &&\\ \midrule 
      \text{ Holographic CFT}  & \left(\dfrac{4\pi}{d}\right)^{d-1} \dfrac{w_{d-1} \ell^{d-1}}{4d G_N} & \left(\dfrac{4\pi}{d}\right)^{d-2} \dfrac{  4(d-2) w_{d-1} \ell^{d-1}}{e^2}  \\[10pt] \bottomrule
    \end{array}
    \]
\end{threeparttable}
    \caption{The coefficients $a$ and $b$ in equation \eqref{eqn:addef} for a variety of CFTs. For the
    free scalar, the results are for $d>3$. $w_{d-1}$ is the area of the unit $(d-1)$-sphere.}
    \label{tab:coefficients}
\end{table}

The organization of this paper is as follows. In Section \ref{sec:CFTonS1}, we give a general argument for the large $T$ behavior in equation (\ref{twistedexpansion}) using  the spurion analysis for the theory obtained by dimensional reduction of the
CFT on the thermal circle. 
In Section 3, we expand the right-hand side of equation (\ref{twistedexpansion}) in characters of representations of $G$ and derive equation (\ref{CharacterExpansion}). In Sections 4 - 6, we discuss examples. 
In Section \ref{sec:U1sym}, we derive the large $T$ behavior when $G=U(1)$ for free field theories and holographic CFTs. In Section \ref{sec:NAsym}, we generalize these results to a non-abelian group $G$. The holographic dual in this case involves the Yang--Mills theory with gauge group $G$, and we need to consider two types of black hole solutions: those with and without non-abelian hair. We show that the two solutions converge at high temperature
and reproduce the behavior in equation  (\ref{CharacterExpansion}). In Section \ref{sec:stabilityRNNA}, we discuss the theormodynamical stability of the black hole with non-abelian hair.

\section{Spurion analysis}
\label{sec:CFTonS1}

Consider a $d$-dimensional CFT on a $(d-1)$-dimensional compact Cauchy surface 
$\Sigma_{d-1}$ times the thermal circle $S^1_\beta$ at temperature $T = 1/\beta$.
We assume that the CFT is invariant under a compact Lie group $G$. To calculate the twisted partition function (\ref{twistedbyg}), we use the approach of \cite{Banerjee:2012iz,Jensen:2012kj,DiPietro:2014bca} and couple the CFT to a background gauge field $A$ with gauge group 
$G$.\footnote{We thank David Simmons-Duffin for discussion on this approach.} Upon dimensional reduction on $S^1_\beta$, dynamical degrees of freedom acquire thermal masses.
The low energy theory on $\Sigma_{d-1}$ is then described by a gauge field $a =A_{|\Sigma_{d-1}}$ coupled to a scalar field $\phi$ in the adjoint representation of $G$, which is related to the holonomy of the gauge field around the thermal circle as
\begin{equation}
    g = \exp\left(i \oint_{S^1_\beta} A\right) \equiv e^{i \phi} \,. 
\end{equation}
The low energy effective Lagrangian in $(d-1)$ dimensions has the derivative expansion, 
\begin{equation}
    \mathcal{L} = {\rm tr}_{\rm Adj} \, \left[ \, T^{d-1} \, V(e^{i \phi}) + 
    c \, T^{d-3} (\mathcal{D}\phi)^2 +g_{YM}^{-2} F^2 + 
    \cdots \, \right] \, ,
    \label{Lagrangian}
\end{equation}
where the scalar potential ${\rm tr}_{\rm Adj} \,  V(g)$ is a class function of $g$ as required by gauge invariance in $d$ dimensions, $\mathcal{D}$ is the covariant derivative, $F= da + a^2$, and $\cdots$ are terms suppressed by $1/T$. The expansion may also include Chern-Simons terms. 
The twisted partition function $Z(T, g)$ is obtained by
setting $g=e^{i\phi}$ to be constant and $a=0$. Therefore, its $g$-dependence 
is captured by the
potential term ${\rm tr}_{\rm Adj} \, V(g)$ in the effective Lagrangian as
\begin{equation}
   Z(T, g)/Z(T, 1)  = \exp\left( - {\rm tr}_{\rm Adj} \,  \left[ T^{d-1} \,V(g) \, {\rm vol} (\Sigma_{d-1}) \right] + \cdots \right).
   \label{twistedenergy}
\end{equation}

Now, we relate the potential ${\rm tr}_{\rm Adj} \,  V(g)$ to the tension of the domain wall which 
generates the $g$-twisted sector in the CFT Hilbert space. To do so, we note
that the Lagrangian density (\ref{Lagrangian})  is of the same form for any smooth
compact manifold $\Sigma_{d-1}$, provided we use the metric of $\Sigma_{d-1}$ to write ${\cal L}$ in a diffeomorphism invariant way. 
In particular, we can choose
$\Sigma_{d-1} = {\tilde S}^1 \times \Sigma_{d-2}$, with ${\tilde S}^1$ having 
unit circumference and
the thermal boundary condition,  and compute $V(g)$ for this geometry. 
By exchanging the thermal circle $S_\beta^1$ with ${\tilde S}^1$ as done for example
in \cite{Shaghoulian:2015kta}, we can 
interpret the twisted partition function $Z(T, g)$ as the untwisted partition 
function in the $g$-twisted sector on $S_\beta^1 \times \Sigma_{d-2}$ with the 
twist along the $S_\beta^1$ direction. 
Since we are computing the partition function
of the CFT, we can rescale the spacetime so that the thermal circle $S_\beta^1$ has
unit circumference and the volume of ${\tilde S}^1 \times \Sigma_{d-2}$ is proportional to $T^{d-1}$. In the limit of $T \rightarrow \infty$, the exponent ${\rm tr}_{\rm Adj} \,  \left[ T^{d-1} \,V(g) \, {\rm vol} (\Sigma_{d-1}) \right] + \cdots $ of equation (\ref{twistedenergy}) can be interpreted as the
ground state energy of the $g$-twisted sector 
on $S_\beta^1 \times \Sigma_{d-2}$
times the circumference $T$ of the rescaled ${\tilde S}^1$.

Since we expect that the ground state energy of the
$g$-twisted sector with $g\neq 1$ is higher than that of the untwisted ground
state, ${\rm tr}_{\rm Adj} \,  V(g)$ should have the global minimum at $g=1$.
Therefore, in the high temperature limit, 
\begin{equation}
      Z(T, g)/Z(T, 1)  \rightarrow C(T) \, \delta(g, 1), ~~~ T \rightarrow \infty,
\end{equation}
for some $C(T)$, 
where $\delta(g,1)$ is the delta-function on the group manifold $G$ localized 
at $g=1$. The delta-function can be expanded in terms of characters as
\begin{equation}
\delta(g, 1) = \sum_R {\dim R} \cdot \chi_R(g),
\end{equation}
where the sum is over unitary irreducible representations of $G$ and the volume of $G$ is normalized to be $1$. Therefore, 
the probability $P_R$ for a random state to be in 
the representation $R$ is proportional to $(\dim R)^2$, for fixed $R$ in the limit of 
$T \rightarrow \infty$. This explains the $(\dim R)^2$ factor in 
equation (\ref{eqn:holographic}).

To reproduce the $\exp\left[ -c_2(R)/(b\,T^{d-1})\right]$ factor 
in equation (\ref{eqn:holographic}), we expand the potential $\Tr V(g)$ around
$g=1$. Since it is a class function of $g$, the expansion should take the form,
\begin{equation}
    {\rm tr}_{\rm Adj} \,  \left[ T^{d-1} \,V(g=e^{i\phi}) \, {\rm vol} (\Sigma_{d-1}) \right]
    = {\rm constant} + \frac{b}{4} \, T^{d-1} \langle \phi, \phi \rangle + \cdots \,.
\end{equation}
The coefficient $b$ must be non-negative since the minimum of ${\rm tr}_{\rm Adj} \,  V(g)$ is at $g=1$.
This reproduces 
equation (\ref{twistedexpansion}). As we will show in the next section, 
this is equivalent to equation (\ref{CharacterExpansion}) and therefore to  equation (\ref{eqn:holographic}).

\section{Expansion in characters}
We have shown that the twisted partition function has the universal high temperature behavior, 
\begin{equation}
    Z(T, g= e^{i \phi})/Z(T, 1) = \exp\left[ - \dfrac{b}{4} T^{d-1} \langle \phi, \phi \rangle + \cdots \right] .
    \label{hightemp0}
\end{equation}
Since it is a class function of $g$,  we can expand it in characters $\chi_R(g)$. The purpose of this section is to 
find the expansion coefficients and derive equation (\ref{CharacterExpansion}).

To do so, we use the fact that the left-hand side of (\ref{hightemp0}) approximately solves the heat equation for $T \gg 1$ as
\begin{equation}
\left( \frac{b T^d}{d-1} \frac{\partial}{\partial T} + \Delta \right)
\left[ \left( \frac{bT^{d-1}}{4 \pi }\right)^{\dim G/2} \ \frac{Z(T, g)}{Z(T, 1)} \right] \simeq 0,
\label{heatequation}
\end{equation}
and obeys the initial condition,
\begin{equation}
   \left.\left( \frac{bT^{d-1}}{4 \pi} \right)^{\dim G/2} \ 
   \frac{Z(T, g)}{Z(T, 1)}\right|_{T = \infty} =
   \delta(g, 1) \, .
   \label{initial}
\end{equation}
Here $\Delta$  is the Laplace operator on the group manifold $G$.
Since each character is an eigenstate of the Laplace operator, 
\begin{equation}
    \Delta \chi_R (g) = - c_2(R) \chi_R(g),
\end{equation}
and since characters make an orthonormal basis of class functions, $\{ \chi_R (g) e^{ - c_2(R)/(b T^{d-1})} \}_R$ gives the complete set of solutions to the heat equation. Therefore, we can expand
\begin{equation}
     \left( \frac{bT^{d-1}}{4 \pi} \right)^{\dim G/2} \ 
   \frac{Z(T, g)}{Z(T, 1)}  \simeq \sum_R \, d_R  \, \chi_R(g) \, \exp\left(- \frac{c_2(R)}{b \, T^{d-1}} \right).
\end{equation}
To determine the expansion coefficient $d_R$, we use the initial condition (\ref{initial}), which can be written as
\begin{equation}
    \sum_R \, d_R  \, \chi_R(g) = \delta(g, 1).
\end{equation}
Since $\delta(g, 1) =  \sum_R \dim R  \cdot  \chi_R(g)$, the expansion coefficients are determined as
\begin{equation}
    d_R =  \dim R \, , 
\end{equation}
and we obtain
\begin{equation}
    Z(T, g)/Z(T, 1) = \left(\frac{4\pi }{bT^{d-1}}\right)^{\dim G/2}
    \sum_R \ \dim R \cdot \chi_R(g)\ \exp\left(- \frac{c_2(R)}{b \, T^{d-1}} + \cdots \right).
\end{equation}

\section{Examples 1: $U(1)$ symmetry} \label{sec:U1sym}

In the remainder of the paper, we will study free field theories and holographic CFTs on 
$S^1_\beta \times S^{d-1}$ and calculate the coefficient $b$ explicitly. The circumference of the thermal circle $S^1_\beta$ is $\beta$, and the radius of the Cauchy surface 
$S^{d-1}$ is normalized to be $1$. 

\bigskip

We begin by studying CFTs with $G = U(1)$. Each state in the Hilbert space can be labeled by a charge $Q$, and the conjectured formula takes the form,
\begin{equation}
    P_Q = \sqrt{\frac{4\pi b}{T^{d-1}}}\, \exp\left[ - \frac{Q^2}{b \ T^{d-1}} \left(1 + \mathcal{O}\left( \frac{1}{T}, \frac{Q^2}{T^{2d-4}}\right)\right)\right]  \,.
\label{U1}
\end{equation}
We verify this by calculating the grand canonical partition function with an imaginary chemical potential $\mu = iT \theta$,
\begin{equation}
    Z(T, \mu = iT\theta) =\Tr\left[ e^{-\beta \hat{H} + i \theta \hat{Q}} \right] \,.
    \label{grandcanonical}
\end{equation}
We assume that $\widehat{Q}$ is quantized in such a way that the field with the smallest non-zero $U(1)$ charge has charge $1$. In the limit of large $T$ and small $\mu$, we show
\begin{equation}
    Z(T, \mu) =\exp\left[a T^{d-1}\left( 1 + \mathcal{O}\left(\frac{1}{T}\right)\right) + \frac{b}{4}\,  {T^{d-3}\mu^2}\left(1+\mathcal{O}\left(\mu^2,\frac{1}{T}\right)\right)\right]\,,
\label{eqn:Z(T,th)}
\end{equation}
for some constants $a$ and $b$. The Fourier transformation of this formula with respect to $\theta = - i \beta \mu$ gives the canonical partition function, which leads to equation \eqref{U1}.

\subsection{Free field theory}\label{sec:U1CFT}

\noindent
{\bf Free scalar theories:}

Consider a massless complex free scalar field $\phi$ in $d$ spacetime dimensions.\footnote{We generally assume that $d > 3$ due to certain subtleties with massless scalar fields in two and three dimensions which we discuss later.} We normalize the $U(1)$ generator $\widehat{Q}$ such that $\phi$ has charge $1$. For such a theory on $\mathbb{R} \times S^{d-1}$, the grand canonical partition function with an imaginary chemical potential is given by \cite{Melia:2020pzd},
 \begin{equation}
    Z_{\text{scalar}}(T,\mu=i T \theta)=
    \exp \left[
    \sum^\infty_{n=1}\frac{e^{-n\beta\frac{d-2}{2}}}{n}
    \text{cos}( n \theta )
    \frac{(1-e^{-2n\beta})}{(1-e^{-n\beta})^d}\right] \,.
    \label{eqn:ZofU1}
\end{equation}
As we are interested in the high temperature limit, that is, when $\theta= - i \beta \mu$ is small, we first expand the exponent in powers of $\theta$ as
\begin{align}
\label{eqn:Zintheta}
    Z_{\text{scalar}}(T,\mu)=\exp\left[\sum_{k=0}^{\infty} C_{k}\theta^{2k}\right] \,,
\end{align}
where the coefficient $C_k$ is given by
\begin{equation}
    C_k=\frac{(-1)^k}{(2k)!} \sum^\infty_{n=1} n^{2k-1} e^{-\frac{(d-2)}{2}n\beta} \frac{(1-e^{-2n\beta})}{(1-e^{-n\beta})^d}\,.
    \label{eqn:Akoriginal}
\end{equation}

At high temperature, one might think that the sum over $n$ in equation \eqref{eqn:Akoriginal} can be approximated by an integral over $x=n\beta$ as
\begin{align}
    C_k\approx \frac{(-1)^k}{(2k)!} T^{2k}\int_0^{\infty}f(x)dx,\quad f(x)=x^{2k-1}e^{-\frac{d-2}{2}x}\frac{(1-e^{-2x})}{(1-e^{-x})^d}\,.
\end{align}
However, we need to be careful when $2k\leq d-1$ as $f(x)$ is singular at $x=0$ and the integral
approximation will fail when $x$ is small. To take this into account, we introduce a cutoff at some small value $x_0$ and use the integral approximation only for $x>x_0$. The terms in the summation in equation \eqref{eqn:Akoriginal} are not converted to integral form when $n$ is such that $n\beta < x_0$. Taking this singular behavior into account, the correct approximation is 
\begin{align}
    C_k\approx\frac{(-1)^k}{(2k)!}\left( 2T^{d-1}\sum_{n=1}^{x_0T}n^{2k-d}
    + T^{2k}\int_{x_0}^{\infty} dx f(x) \right) \,.
    \label{eqn:approxcoeff}
\end{align}
It is straightforward to show that equation \eqref{eqn:approxcoeff} is independent of $x_0$ for large values of $T$. In this way, we find that the coefficients $C_k$ can be approximated as
\begin{subequations}
\begin{numcases}{C_k\approx}
    \scalemath{.91}{\frac{(-1)^k}{(2k)!} \left( \int_0^{\infty}dx\ x^{2k-1}e^{-\frac{d-2}{2}x} \frac{(1-e^{-2x})}{(1-e^{-x})^d} \right)T^{2k}} & $2k\geq d$,
    \label{eqn:2kGREdminus1}\\
    \scalemath{.91}{\frac{(-1)^k}{(2k)!} \left( 2\log T+2\gamma+2\log x_0 +\int_{x_0}^{\infty}dx\ x^{2k-1}e^{-\frac{d-2}{2}x} \frac{(1-e^{-2x})}{(1-e^{-x})^d} \right)T^{d-1}} & $2k=d-1$, \label{eqn:2kISdminus1} \\
    \begin{aligned}
    &\scalemath{.91}{\frac{(-1)^k}{(2k)!} 2\zeta(d-2k)T^{d-1}} \\ 
    &\qquad\qquad\scalemath{.91}{-\frac{(-1)^k}{(2k)!}T^{2k} \left( \frac{2x_0^{2k-d+1}}{d-2k-1} + \int_{x_0}^{\infty}dx\ x^{2k-1}e^{-\frac{d-2}{2}x} \frac{(1-e^{-2x})}{(1-e^{-x})^d} \right)}
    \end{aligned} & $2k<d-1$. \label{eqn:2kLESdminus1}
\end{numcases}
\end{subequations}
The constant $\gamma$ appearing in equation \eqref{eqn:2kISdminus1} is the Euler--Mascheroni constant. At $\theta=0$, the partition function is $Z_{\text{scalar}}(T,0) = e^{C_0}$. Since equation \eqref{eqn:2kLESdminus1} gives $C_0 = 2\zeta(d) \, T^{d-1}$,
the coefficient $a$ in equation \eqref{eqn:addef} is given by $a=2\zeta(d)$ for the massless free scalar.

For $d>3$, equation \eqref{eqn:2kLESdminus1} gives
\begin{equation}
    C_1\approx \zeta(d-2)\, T^{d-1} \,,
    \label{eqn:A1}
\end{equation}
where we ignore the second term of equation \eqref{eqn:2kLESdminus1}, since it is subleading in $1/T$. Thus we find that the grand canonical partition function is
\begin{equation}
    Z_{\text{scalar}}(T,\mu) \approx \exp\left[ \zeta(d-2) \ T^{d-3} \mu^2(1+\mathcal{O}(\mu^2,1/T)) \right]  Z_{\text{scalar}}(T,0) \,.
\end{equation}
In summary, the grand canonical partition function of the massless free complex scalar field theory in $d > 3$ demonstrates the universal behavior at high temperature as in equation \eqref{eqn:Z(T,th)}, with constants
\begin{equation}
    a = 2\zeta(d) \,, \qquad b = 4\zeta(d-2) \,.
\end{equation} 

When $d=3$, there is a logarithmic term in the exponent due to equation \eqref{eqn:2kISdminus1},
\begin{equation}
    Z_{\text{scalar}}(T,\mu) \approx \exp\left[ (\log T+2.96351...)\mu^2(1+\mathcal{O}(\mu^2,1/T))\right] Z_{\text{scalar}}(T,0) \,.
    \label{eqn:3dscalar}
\end{equation}
However, the massless scalar field at $d=3$ does not make sense at finite temperature since it has the same infrared issue as that of the massless scalar field at $d=2$. We believe that the appearance of the $\log T$ singularity is a reflection of the infrared pathology in this case.

\bigskip

\noindent
{\bf Free spinor theories:}

For the massless scalar field, we cannot consider theories in $d=2,3$ due to the infrared problem. As it is good to also have an example in these dimensions, we consider the theory of a free spinor field. 

In two dimensions, the grand canonical partition function of a free complex Weyl spinor is given by
\begin{equation}
    Z_{\text{spinor}}(T,\mu= i T \theta) =
    \prod_{n=1}^{\infty}
    (1+e^{-\beta(n-\frac{1}{2})}e^{i\theta})(1+e^{-\beta(n-\frac{1}{2})}e^{-i\theta}) \,.
\end{equation}
We can transform this into the plethystic form as
\begin{align}
\begin{aligned}
    Z_{\text{spinor}}(T,\mu=iT\theta) &=\exp \left[ \sum_{n=1}^{\infty} \left( \log (1+e^{-\beta(n-\frac{1}{2})}e^{i\theta}) +\log (1+e^{-\beta(n-\frac{1}{2})}e^{-i\theta}) \right)\right] \\
    &=\exp \left[ -\sum_{n,m=1}^{\infty}\frac{(-1)^m}{m}\left( e^{-\beta m(n-\frac{1}{2})}e^{im\theta}+e^{-\beta m(n-\frac{1}{2})}e^{-im\theta}\right) \right] \\ 
    &=\exp \left[ -\sum_{m=1}^{\infty}\frac{(-1)^m}{m} \frac{\operatorname{cos}(m\theta)}{\operatorname{sinh}\frac{m\beta}{2}} \right] \,.
\end{aligned}
\end{align}
As in the free scalar case, we expand the exponent of the partition function in $\theta$ as
\begin{align}
\label{eqn:Zfintheta}
    Z_{\text{spinor}}(T,i\mu)=\exp\left[\sum_{k=0}^{\infty} D_k\theta^{2k}\right]\,,
\end{align}
for some coefficients $D_k$. We find that
\begin{equation}
    D_1
    =-\frac{1}{2}\sum_{m=1}^{\infty}(-1)^m \frac{m}{\operatorname{sinh}\frac{m\beta}{2}}
    =-\frac{1}{2}\sum_{n=1}^{\infty}\left( \frac{2n}{\operatorname{sinh}(n\beta)} -\frac{2n-1}{\operatorname{sinh}\left(\frac{2n-1}{2}\beta\right)}
    \right) \,,
\end{equation}
where we split the series into $m=2n$ and $m=2n-1$ terms and sum them as pairs, which is valid as $D_1$ converges due to the hyperbolic sine function in the denominator. At high temperature, we can approximate the summation as an integration over $x=n\beta$:
\begin{equation}
    D_1\approx
    -\frac{\text{T}^2}{4}\int_{0}^{\infty}dx(f(x+\beta)-f(x)) \approx -\frac{\text{T}^2}{4}\int_{0}^{\infty}dx f'(x)\beta \approx \frac{T}{4}, \quad
    f(x)=\frac{x}{\operatorname{sinh}x}\,.
    \label{eqn:distwoAone}
\end{equation}
Similarly,
\begin{equation}
    D_0 = -\sum_{m=1}^{\infty}\frac{(-1)^m}{m\,\operatorname{sinh}\frac{m\beta}{2}} \,.
\end{equation}
In this case, we need  the cutoff $x_0$ to covert the sum into an integral as
\begin{equation}\label{eqn:distwoAzero}
    D_0\approx
    -2T\sum_{m=1}^{x_0T}\frac{(-1)^m}{m^2} - \frac{T}{2}\int_{x_0}^{\infty}dx g'(x)\approx \zeta(2)T, \quad g(x)=\frac{1}{x \ \operatorname{sinh}{\frac{x}{2}}} \,,
\end{equation}
where we used the zeta function identity $-\sum_{m=1}^{\infty}\frac{(-1)^m}{m^2}=\frac{\zeta(2)}{2}$. 

Let us turn to $d=3$, where the grand canonical partition function of the free spinor theory is
\begin{align}
    Z_{\text{spinor}}(T,\mu= iT\theta) &= \prod_{n=0}^{\infty} (1+e^{-\beta(n+1)}e^{i\theta})^{2n+1}(1+e^{-\beta(n+1)}e^{-i\theta})^{2n+1} \\ 
    &=\exp \left[ -\sum_{m=1}^{\infty}\frac{(-1)^m}{m} e^{-\frac{m\beta}{2}}\frac{\operatorname{coth}\frac{m\beta}{2}}{\operatorname{sinh}\frac{m\beta}{2}}\operatorname{cos}(m\theta) \right] \,.
\end{align}
Expanding the exponent in powers of $\theta$, we find the coefficients to be
\begin{align}\label{eqn:disthreeA}
\begin{aligned}
    D_0&\approx -\sum_{m=1}^{x_0T}(-1)^m\frac{4}{m^3\beta^2} + \frac{T}{2}\int_{x_0}^{\infty}dxf'(x)\approx 3\zeta(3),\quad
    f(x)=\frac{1}{x}e^{-\frac{x}{2}}\frac{\operatorname{coth}\frac{x}{2}}{\operatorname{sinh}\frac{x}{2}},\\
    D_1&\approx -\sum_{m=1}^{x_0T}(-1)^m\frac{4}{m\beta^2} + \frac{T}{2}\int_{x_0}^{\infty}dxg'(x)\approx 4T^2\log{2},\quad
    g(x)=xe^{-\frac{x}{2}}\frac{\operatorname{coth}\frac{x}{2}}{\operatorname{sinh}\frac{x}{2}}\,,
\end{aligned}
\end{align}
where we used zeta function identities 
$\sum_{m=1}^\infty\frac{(-1)^m}{m}=\log 2$ and $\sum_{m=1}^\infty\frac{(-1)^m}{m^3}=-\frac{3}{4}\zeta(3)$.

Combining these results, we find
\begin{subequations}
\begin{numcases}{Z_{\text{spinor}}(T,\mu)\approx}
    \exp\left[\frac{1}{4T}\mu^2(1+O(\mu^2,1/T)) \right] Z_{\text{spinor}}(T,0)  & $d=2$ \,, \label{eqn:apxZfTthdIS2}\\
    \exp\left[4\log 2\,\mu^2(1+O(\mu^2,1/T))\right] Z_{\text{spinor}}(T,0) & $d=3$ \,,
    \label{eqn:apxZfTthdIS3}
    \end{numcases}
where the partition functions at $\beta\mu=0$ for both dimensions are given by,
\begin{numcases}{Z_{\text{spinor}}(T)=Z_{\text{spinor}}(T,0)\approx}
    e^{\zeta(2)T}  & $d=2$, \label{eq:spinord2} \\
    e^{3\zeta(3)T^2}  & $d=3$ \,. \label{eq:spinord3}
\end{numcases}
\end{subequations}
Equations \eqref{eq:spinord2} and \eqref{eq:spinord3} show that the free Weyl spinor theory also demonstrates the universal behavior in equation \eqref{eqn:Z(T,th)} at high tempearture with the coefficients
\begin{equation}
a = \zeta(2) = \frac{\pi^2}{6} \,, \qquad b=1 \,,
\end{equation}
for $d=2$, and
\begin{equation}
    a = 3\zeta(3)\,, \qquad b = 16 \log 2 \,,
\end{equation}
for $d=3$. 
For $d=2$, the Cardy formula gives $a = \pi^2 (c_L+c_R)/6$, and the above value of $a$ is
consistent with $(c_L, c_R) = (1, 0)$ for the complex Weyl spinor. 
As expected, the result at $d=3$ is free from the $\log T$ singularity we saw for the free scalar field in equation (\ref{eqn:3dscalar}).

\subsection{Holographic CFT}\label{sec:U1hol}

We now consider a holographic CFT, whose bulk theory is described at low energy in terms of the Einstein gravity coupled to the Maxwell field and a finite number of matter fields in AdS$_{d+1}$. The action
of the theory is given by
\begin{equation}
    I = \int d^{d+1}x\sqrt{-g}\left[\frac{1}{16\pi G_N}\left(R + \frac{d(d-1)}{\ell^2}\right) - \frac{1}{4e^2} F^2  + \cdots \right] \,,
\label{eqn:EMaxaction}
\end{equation}
where $\cdots$ represents matter field terms. The curvature radius $\ell$ is related to the cosmological constant as $\Lambda=-d(d-1)/2\ell^2$. To calculate the grand canonical partition function, we impose the boundary condition that the boundary geometry is $S^1_\beta \times S^{d-1}$ and the gauge field $A$ has the holonomy around the thermal circle $S^1_\beta$ at the boundary given by
\begin{equation}
    \exp\left( i \oint_{S^1_\beta} A_\tau \right) = e^{\beta\mu} \,,
\end{equation}
where $\mu$ is identified with the chemical potential of the boundary CFT.
We solve the Einstein and Maxwell equations assuming the spherical symmetry on $S^{d-1}$ and setting all other matter fields to zero.

There are two classical solutions under these conditions; one is the thermal AdS and the other is the AdS Reissner--Nordstrom (RN) black hole. 
At high temperature, the RN solution is dominant \cite{Chamblin:1999tk,Chamblin:1999hg}. The RN solution can be written in static coordinates as
\begin{subequations}
\begin{align}
    & ds^2 = V(r)d\tau^2+\frac{dr^2}{V(r)}+r^2d\Omega^2_{d-1},\quad
    V(r)=1-\frac{m}{r^{d-2}}+\frac{v q^2}{r^{2d-4}}+\frac{r^2}{\ell^2},\\
    & A =-i\sqrt{\frac{d-1}{2(d-2)}}\left(\frac{q}{r_H^{d-2}}-\frac{q}{r^{d-2}}\right) d\tau,  \quad v =\frac{4\pi G_N}{e^2} \, , 
    \label{eqn:U(1)RNA}
\end{align}
\end{subequations}
where $m$ and $q$ are related to the ADM mass and the charge of the black hole \cite{Romans:1991nq,london1995arbitrary,Chamblin:1999tk}.
This solution has its ADM mass, charge, temperature and entropy given by,
\begin{subequations}
\begin{align}
\label{eqn:ddimRNTherm}
    M & =\frac{(d-1)w_{d-1}}{16\pi G_N} r_H^{d-2}  \left( 1+\frac{v q^2}{r_H^{2d-4}}+\frac{r_H^2}{l^2} \right),\\
    Q & =\sqrt{2(d-1)(d-2)}\left(\frac{w_{d-1}}{8\pi G_N}\right)v q,\\
    T &=\frac{d-2}{4\pi r_H}  \left( 1-\frac{v q^2}{r_H^{2d-4}}\right)+\frac{r_H d}{4\pi l^2},\\
    S &=\frac{w_{d-1}}{4G_N}r_H^{d-1},
\end{align}
\end{subequations}
where $w_{d-1}$ is the surface area of the unit ($d-1$)-sphere, and the horizon radius $r_H$ is the largest real positive root of $V(r)$ \cite{Chamblin:1999hg, arnowitt1959dynamical,abbott1982stability,hawking1996gravitational}. 
The chemical potential of the black hole system is related to the charge $Q$ as
\begin{equation}
    \mu = \sqrt{\frac{d-1}{2(d-2)}}\ \frac{q}{r_H^{d-2}}
    = \frac{e^2}{(d-2)w_{d-1}}\frac{Q}{r_H^{d-2}}.
\label{eqn:chempotfromA}
\end{equation}
By the AdS/CFT correspondence, the grand canonical partition function of the CFT can be
calculated using the Euclidean action for this solution.

At high temperature, the horizon radius $r_H$ of the stable black hole grows linearly in 
the temperature as
\begin{equation}
    T\approx\frac{r_H d}{4\pi \ell^2} \left( 1-X\right),\quad X=\frac{d-2}{d}\frac{v q^2 \ell^2}{r_H^{2d-2}}\,,
\end{equation}
where we keep $X$ as small which is equivalent to small $|\mu|$ approximation in free field calculation.
 The grand potential $\Phi(T,\mu)$ is related to the grant canonical partition function as
\begin{equation}
    Z_{AdS}(T,\mu)=e^{-\beta \Phi(T,\mu)},
\end{equation}
and is given by the Euclidean action of the RN solution,
\begin{align}
    \Phi(T,\mu)=M-TS-\mu Q\approx -\frac{w_{d-1}r_H^d}{16\pi G_N\ell^2} \left( 1 + \frac{d}{d-2}X\right).
\end{align}
Using
\begin{align}
    r_H=\frac{4\pi \ell^2}{d}\frac{T}{1-X},\quad 
    X = \frac{d(d-2)^2}{8\pi^2\ell^2(d-1)}\frac{v \mu^2}{T^2}+O(X^2),
\label{eqn:rHXTQdep}
\end{align}
we find
\begin{align}
\begin{aligned}
    -\beta \Phi(T,\mu) \approx
    \frac{w_{d-1}(4\pi \ell^2/d)^d}{16\pi G_N \ell^2}T^{d-1} + \frac{w_{d-1}(d-2)}{e^2 }\left(\frac{4\pi \ell^2}{d}\right)^{d-2}\mu^2 T^{d-3}.
\label{eqn:U1RNGibbs}
\end{aligned}
\end{align}
Rescaling the temperature as $\ell T \rightarrow T$, 
the grand canonical partition function of the dual CFT on the sphere with unit radius is given by
\begin{equation}
  Z_{CFT}(T,\mu)
    \approx \exp \left[ w_{d-1}\left( \dfrac{4 \pi}{d} \right)^{d-1}\left(
    \dfrac{\ell^{d-1}}{4dG_N}T^{d-1} + \dfrac{d(d-2)\ell^{d-1}}{4\pi e^2}\mu^2T^{d-3}\right)
    \right].
    \label{U1grandcanonical}
\end{equation}
This determines the coefficients $a$ and $b$ of equation \eqref{eqn:Z(T,th)} in this case as
\begin{align}
    a=\left(\dfrac{4\pi}{d}\right)^{d-1} \dfrac{w_{d-1} \ell^{d-1}}{4d G_N} , \quad
    b=\left(\dfrac{4\pi}{d}\right)^{d-2} \dfrac{ 4(d-2) w_{d-1} \ell^{d-1}}{e^2}.  
\end{align}

\section{Examples 2: non-abelian symmetry} \label{sec:NAsym}
When $G$ is 
non-abelian, we utlize the fact that the twisted partition function $Z(T, g)$ is a class function 
invariant under the conjugation $g \rightarrow h g h^{-1}$ for any $h$. This allows us to restrict $g$ 
to the maximum torus of $G$ and simplify our calculation.
In both free field theories and holographic CFTs, we find
\begin{equation}
    Z(T, g= e^{i\phi}) = \exp\left[ - \dfrac{b}{4} T^{d-1} \langle \phi, \phi \rangle + \cdots \right] Z(T, g=1),
    \label{non-Abelian}
\end{equation}
where the constant $b$ is independent of $\phi$ or $T$.

\subsection{Massless free scalar}\label{sec:GCFT}

Suppose a compact Lie group $G$ has a faithful 
unitary representation $\rho$ with $\dim \rho = n$.
Consider $n$ massless scalar fields in $d$ dimensions. Though the theory has a larger symmetry of $O(n)$, we focus on its $G$ subgroup. 
We would like to calculate the finite temperature partition function of this theory with an insertion of $g \in G$ as
\begin{equation}
    Z(T, g) = \Tr\left[U(g) e^{-\beta \widehat{H}}\right].
\end{equation}

Since $Z(T,g)$ is a class function of $g$,  without loss of generality, $g$ can be restricted to the maxim torus of $G$. In this case, $U(g)$ acts as a multiplication of a phase factor on each of the scalar fields. We can then apply equation (\ref{eqn:ZofU1}) for $G= U(1)$ to each scalar field and assemble the results to obtain
 \begin{equation}
    Z_{\text{scalar}}(T,g)=
    \exp \left[
    \sum^\infty_{n=1}\frac{e^{-\frac{d-2}{2}n\beta}}{n}
   \frac{\chi_\rho(g^n)+\chi_\rho^*(g^n)}{2}
    \frac{(1-e^{-2n\beta})}{(1-e^{-n\beta})^d}\right],
    \label{eqn:ZofG}
\end{equation}
where $\chi_\rho$ is the character of the representation $\rho$ and $\chi_\rho^*$ is that for its conjugate. 
Writing $g = e^{i\phi}$ and expanding in powers of $\phi$, 
\begin{align}
     \frac{\chi_\rho(g^n)+\chi_\rho^*(g^n)}{2} &= {\rm tr}_\rho \left( 1 - n^2 \phi^2 + \cdots \right) \nonumber \\
     &= \dim \rho - \dfrac{\dim \rho}{\dim G} c_2(\rho) \langle \phi, \phi \rangle n^2 + \cdots ,
\end{align}
where ${\rm tr}_\rho$ is the trace over the representation $\rho$ and $\langle \phi, \phi \rangle 
= {\rm tr}_{\rm Adj} \,  \phi^2$.
 We can repeat the calculation of $G= U(1)$ in Section \ref{sec:U1sym} to obtain
\begin{equation}
    Z_{\text{scalar}}(T,e^{i\phi} )\approx
    \exp\left[ -\zeta(d-2) \ T^{d-1}\frac{\dim \rho}{\dim G}c_2(\rho)\langle \phi, \phi \rangle + \cdots \right] Z_{\text{scalar}}(T, g=1) ,
\end{equation}
where we assumed $d > 3$.

\subsection{Holographic CFT}\label{sec:GHol}

Consider a holographic CFT in $d$ dimensions, whose bulk theory is described in low energy in terms of the Einstein gravity coupled to the Yang--Mills field with gauge group $G$ and a finite number of matter fields in AdS$_{d+1}$. The action of the theory is given by
\begin{equation}
    I = \int d^{d+1}x\sqrt{-g}\left[\frac{1}{16\pi G_N}\left(R+\frac{d(d-1)}{l^2}\right) - \frac{1}{4e^2}\langle F,F\rangle + \cdots \right],
\end{equation}
where $F$ is in the Lie algebra of gauge group $G$ and $\cdots$ represents matter field terms. 
To calculate the grand canonical partition function, we impose the boundary condition that the boundary geometry is $S^1_\beta \times S^{d-1}$ and the gauge field $A_\mu$ has the holonomy around the thermal circle $S^1_\beta$ as
\begin{equation}
    \mathcal{P}\exp\left( i\oint_{S^1_\beta} A \right) = e^{\beta \mu} =g.
\label{eqn:boundarycdn}
\end{equation}
We assume that the solution is spherically symmetric on $S^{d-1}$, and  
all the other matter fields are set to zero. We calculate the field strength and the stress-energy tensor as
\begin{equation}
    F_{\mu\nu}=\partial_\mu A_\nu -\partial_\nu A_\mu -i \left[A_\mu,A_\nu\right],\quad
    T_{\mu\nu} = \frac{1}{e^2}\left(\langle F_{\mu\alpha},F_\nu^\alpha\rangle-\frac{1}{4}g_{\mu\nu}\langle F_{\alpha\beta},F^{\alpha\beta}\rangle\right).
\end{equation}

There are three classical solutions under these conditions. The first is the thermal AdS, 
\begin{equation}
    ds^2 = \left(1-\frac{\Lambda r^2}{3}\right)d\tau^2 + \frac{dr^2}{1-\frac{\Lambda r^2}{3}} + r^2 d\Omega^2_{d-1}\,,\quad
    A = -i\mu\, d\tau\,.
\end{equation}
The second  makes use of the $U(1)$ RN solution \eqref{eqn:EMaxaction}, $ds^2_{U(1)}$ and $a_\mu$ with the
chemical potential $\mu_{U(1)}$, by the substitution,
\begin{equation}
    ds^2 = ds_{U(1)}^2,\quad A_\mu = \frac{\phi}{\langle \phi,\phi\rangle^{1/2}} \ a_\mu  \, ,
    \quad \mu = \frac{\phi}{\langle \phi,\phi \rangle^{1/2}} \mu_{U(1)}. 
    \label{abelianansatz}
\end{equation} 
Since $H$ commutes with itself, the Yang--Mills equation for $A_\mu$ reduces to the Maxwell 
equations for $a_\mu$.
The rescaling by $\langle \phi, \phi \rangle^{-1/2}$ is needed to match the stress energy tensors of both systems.

The third is a genuinely non-abelian solution. A dyonic black hole solution with $SU(N)$ hair is known in $AdS_4$ \cite{Shepherd_2016}. Here,
we construct a purely electric black hole solution with $SU(2)$ hair with the following ansatz \cite{Bjoraker:2000qd},
\begin{align}
\begin{aligned}
    &ds^2 = -\mu(r)\sigma(r)^2dt^2 + \frac{dr^2}{\mu(r)}+r^2d\theta^2+r^2\text{sin}^2\theta d\phi^2\,,\quad  \\
    &\mu(r)=1-\frac{2m(r)}{r}-\frac{\Lambda r^2}{3},\\
    &A_\mu = h(r)\frac{\tau_3}{2} dt + w(r)\frac{\tau_1}{2} d\theta +\left( \text{cot}\theta \frac{\tau_3}{2} + w(r) \frac{\tau_2}{2} \right)\text{sin}\theta d\phi\,,
\label{eqn:NAansatz}
\end{aligned}
\end{align}
where we use Pauli matrices $\tau_{1,2,3}$ as generators of the Lie algebra of $SU(2)$ and the inner product
is defined as twice the trace of two elements' multiplication. 
The $AdS$ boundary condition requires $\sigma(r \rightarrow \infty) = 1$.
The functions, $\sigma(r)$, $m(r)$, $h(r)$, and $w(r)$, are determined by numerically solving the Einstein Yang--Mills equations, which take
the form \cite{Bjoraker:2000qd}, 
\begin{subequations}
\begin{align}
    & h'' =h'\left(\frac{\sigma'}{\sigma}-\frac{2}{r}\right)+\frac{2w^2}{\mu r^2}h, \label{EOM1}\\ 
    & w''+w'\left(\frac{\sigma'}{\sigma}+\frac{\mu'}{\mu}\right)+\frac{wh^2}{\sigma^2\mu^2}+\frac{w(1-w^2)}{\mu r^2} =0 \label{EOM2}\\ 
    & m'= v\left(\frac{r^2 h'^2}{2\sigma^2} + \frac{w^2h^2}{\sigma^2\mu}+\mu w'^2 + \frac{1}{2r^2}(1-w^2)^2\right), \label{EOM3}\\ 
    &\ \sigma' = v\left(\frac{2\sigma w'^2}{r} + \frac{2w^2h^2}{\sigma \mu^2 r}\right), \label{EOM4}
\end{align}
\label{eqn:EYMeqns}
\end{subequations}
\hspace{-8pt}
where the prime denotes differentiation with respect to $r$. The horizon radius $r_H$ is defined as the largest solution to $\mu(r)=0$ and $v=4\pi G_N/e^2$.  

Since we have the three possible solutions, 
we should determine which one gives the dominant contribution to the partition function. Above the Hawking--Page temperature, we should consider either the second or the third solution. It turns out that the two solutions converge at high temperature. This is because, as the temperature rises, the horizon grows and approaches the AdS boundary, where the interaction terms in the bulk equations of motion are suppressed. This expectation will be confirmed by the numerical
computation below.

In the asymptotically $AdS$ case, 
there are stable hairy black hole solutions, and those with $SU(N)$ hair have been extensively studied 
\cite{Volkov_1999,Shepherd_2016,Bjoraker:2000qd,Winstanley:2015loa,Volkov:2016ehx}. In particular, Bjoraker and Hosotani in \cite{Bjoraker:2000qd} discussed
the existence of a purely electric $SU(2)$ charged black hole in $AdS_4$, which is of our interest, but it has not been constructed explicitly.

Let us construct the genuinely non-abelian 
solution with $SU(2)$ purely electric hair in $AdS_4$ at high temperature. We determine $\sigma(r)$, $m(r)$, $h(r)$, and $w(r)$
by integrating equations \eqref{eqn:EYMeqns} from the horizon to the infinity.
Since a thermodynamically stable black hole has a large horizon at high temperature, 
we can expand them in the inverse powers of $r_H$ as
\begin{align}
\begin{aligned}
    h(r) &=r_H\wth_0(\wtr) + r_H^{-1}\wth_1(\wtr) + O(r_H^{-2}),\quad
    && m(r)=r_H^3\wtm_0(\wtr)+r_H\wtm_1(\wtr)+O(1),\\
    \sigma(r) &=1+r_H^{-2}\wts_1(\wtr)+O(r_H^{-3}),\quad
    &&\, w(r)=\wtw_0(\wtr)+r_H^{-2}\wtw_1(\wtr)+O(r_H^{-3}),\\
    \mu(r) &=r_H^2\widetilde{\mu}_0+\widetilde{\mu}_1+O(r_H^{-1}),
\label{eqn:fnexpansions}
\end{aligned}
\end{align}
where $\wtr=r/r_H$. Once we substitute this expansion into Einstein Yang--Mills equations \eqref{eqn:EYMeqns}, and solve the leading order equations, we get leading value of the functions:
\begin{align}\label{eqn:hmzerothsol}
    \wth_0(\wtr)=h^{\prime}_H\left(1-\frac{1}{\wtr}\right),\quad
    \wtm_0(\wtr)=\left( \frac{-\Lambda+3v h_H^{\prime\; 2}}{6}-\frac{v h_H^{\prime\; 2}}{2 \wtr} \right),\quad \tilde{\sigma}_0(\tilde{r})=1,
\end{align}
and $\tilde{w}_0$ is the solution of
\begin{equation}
    0 =\ddot{\wtw}_0+\frac{3vh_H^{\prime\; 2}(\wtr-2)-\Lambda\wtr(1+2\wtr^3)}{\wtr(\wtr-1)(-3vh_H^{\prime\; 2}-\Lambda(\wtr+\wtr^2+\wtr^3))}
    \dot{\wtw}_0+\frac{9h_H^{\prime\; 2} \wtr^2}{(3vh_H^{\prime\;2}+\Lambda(\wtr+\wtr^2+\wtr^3))^2}\wtw_0.
    \label{eqn:wEOMdetailed}
\end{equation}
Here, $h^{\prime}_H=\frac{dh(r)}{dr}|_{r=r_H}\sim\frac{d\wth_0(\wtr)}{d\wtr}|_{\wtr=1}$.
Then, the leading thermodynamic quantities of the black hole with non-abelian hair are given by \cite{Bjoraker:2000qd},
\begin{subequations}
\begin{align}
    Q_E &= \frac{4\pi}{e^2}h'(r)r^2\frac{\tau_3}{2}\Big|_{r\to\infty}\, \quad \xrightarrow{r_H\to\infty}\ \  \frac{4\pi}{e^2}r_H^2 h'_H\frac{\tau_3}{2}, \label{eqn:numsolQE}\\
    Q_M &= \frac{4\pi}{e^2}(1-w(r)^2)\frac{\tau_3}{2}\Big|_{r\to\infty}\xrightarrow{r_H\to\infty}\ \  \frac{4\pi}{e^2}\left(1-\wtw_0(\wtr)^2\right)\frac{\tau_3}{2}, \label{eqn:numsolQM}\\
    M &= \frac{m(r)}{G_N}\Big|_{r\to\infty}\qquad\quad\ \ \ \  \xrightarrow{r_H\to\infty}\ \   \frac{-\Lambda+3v h_H^{\prime\; 2}}{6G_N}r_H^3 , \label{eqn:numsolM}\\
    T &= \frac{1}{4\pi}\sigma(r_H)\mu'(r_H)
    \quad\ \ \ \;    \xrightarrow{r_H\to\infty}\ \  \frac{r_H}{4\pi}(-\Lambda-vh_H^{\prime\; 2}), \label{eqn:numsolT}\\
    S &=  \frac{\pi r_H^2}{G_N}.
\end{align}
\label{eqn:QMTSNABH}
\end{subequations}
\hspace{-8pt}
The AdS boundary condition implies
$\wtw_0(\wtr \rightarrow \infty) = 1$. Since it is known that the black hole is
unstable if $w(r)$ has a node (a nontrivial solution to $w(r)=0$) \cite{Volkov_1995,Bjoraker:1999yd},
we require $\wtw_0(\wtr)$ be positive everywhere.
Under these conditions, we find a unique solution for $\wtw_0$ when $\Lambda$, $v$ and $h'_H$ are given.
This establishes the existence of a stable (nodeless) solution in leading order
for given values of  $r_H$, $\Lambda$, $v$, and $h'_H$, provided $h_H^{\prime\; 2},\, v h_H^{\prime\; 2} < -\Lambda$, which are always satisfied for large enough $T$.

\begin{figure}[H]
	\centering
	\begin{subfigure}{0.45\textwidth}
	\centering
	\includegraphics[width=.97\textwidth]{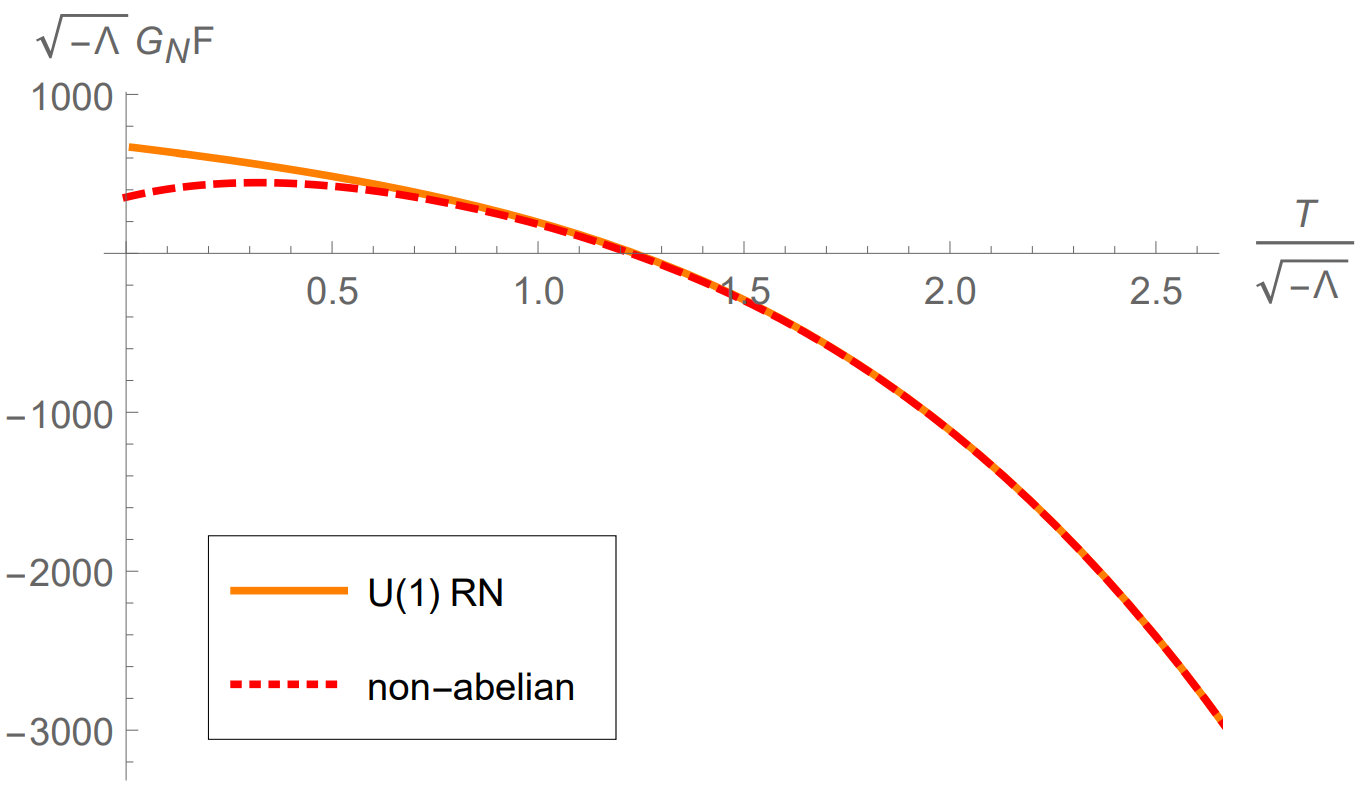}
	\caption{Helmholtz free energies with respect to the temperature}
	\end{subfigure}\qquad
    \begin{subfigure}{0.45\textwidth}
    \centering
	\includegraphics[width=.97\textwidth]{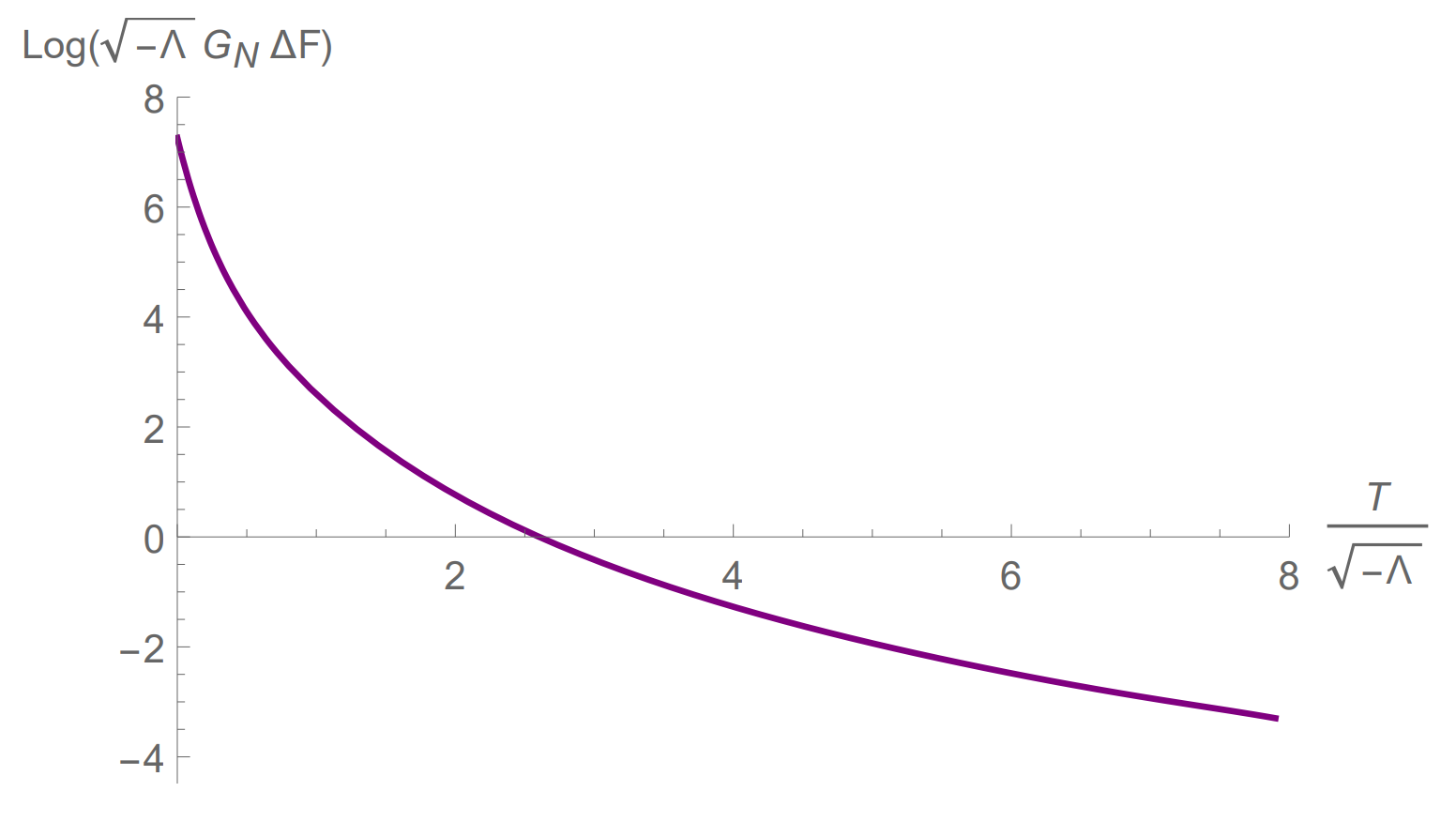}
	\caption{The difference in Helmholtz free energies of the two solutions with respect to the temperature}
	\end{subfigure}
	\caption{They are plotted at a fixed value of $\Lambda=-1,\,v=1$ and $\sqrt{-\Lambda}\,G_NQ=100$.}
	\label{fig:tvsf}
\end{figure}

As expected, at high temperature, 
the thermodynamic quantities of the solution converge to those of the embedded $U(1)$ RN black hole as
\begin{align}
\begin{aligned}
    M=\frac{r_H}{2G_N} \left(-\frac{\Lambda r_H^2}{3}+\frac{e^2\,G_NQ^2}{4\pi r_H^2}\right),\quad &
    Q_E= Q\frac{\tau_3}{2},\\
    T=\frac{1}{4\pi r_H}\left(-\Lambda r_H^2 - \frac{e^2\,G_NQ^2}{4\pi r_H^2} \right),\quad &
    S=\frac{\pi r_H^2}{G_N}.
\end{aligned}
\label{eqn:MQTSembRN}
\end{align}
In Figure \ref{fig:tvsf}, we show the Helmholtz free energies of the two solutions as
functions of $T$, with $Q$, $v$ and $\Lambda$ fixed as
\begin{align}
    \sqrt{-\Lambda}G_NQ=100,\quad v=1,\quad\Lambda=-1.
\end{align}
We observe that Helmholtz free energies of the two solutions converge at high temperature.
We also note that, if we look at smaller temperature, the free energy of the $U(1)$ RN black hole
(in the orange curve)
becomes larger than that of the genuinely non-abelian solution 
(in the dotted red curve). We will discuss more about it in Section \ref{sec:stabilityRNNA}.

Having our expectation confirmed,  we can utilize the $U(1)$ RN solution to estimate
the high temperature behavior of the holographic CFT with non-Abelian
global symmetry $G$ in any dimensions.
In particular,
\begin{equation}
    Z_{G}\left( T, \mu = \frac{\phi}{\langle \phi, \phi\rangle^{1/2}} \mu_{U(1)} \right)  \sim Z_{U(1)}\left(T,\mu_{U(1)} \right), ~~~ T \gg \dfrac{1}{\ell} \, ,
\end{equation}    
where $Z_G$ denotes the grand canonical partition function for the Einstein Yang--Mills
system in $AdS_{d+1}$ with gauge group $G$ and
$Z_{U(1)}$ is that for the Einstein Maxwell system.
By using equation (\ref{U1grandcanonical}), we obtain
\begin{align}
    Z_G(T,\mu) & = \exp\left[a \ T^{d-1} + \frac{b}{4} T^{d-3} \langle \mu, 
    \mu \rangle  + \cdots \right] \nonumber\\
    & = \exp\left[a \ T^{d-1} - \frac{b}{4} T^{d-1} \langle \phi, 
    \phi \rangle  + \cdots \right] ,
\end{align}
where $\beta \mu =i\phi$.

\section{Stability of black hole with non-abelian hair}\label{sec:stabilityRNNA}

In the holographic CFT with non-abelian gauge symmetry, there are two types of black holes solutions, with and without non-abelian hair. In the previous section, we showed that the two
solutions converge at high temperature. 
Since the two solutions differ at lower temperature, 
it is interesting to find out which solution is preferred theormodynamically. In this section, we calculate $1/T$ corrections
to the Helmholtz free energies of the two solutions at the same temperature
and charge. We find that the 
black hole with non-abelian hair has a lower free energy
and is more stable. 
To be specific, we consider $G=SU(2)$ though we believe
the results apply to any compact Lie group. 

Since we know the exact solution without non-abelian hair, we 
focus on evaluating $1/T$ corrections to the solution with hair. 
We start with the equations, 
\begin{align}
 \label{eqn:EYMsubleading}
    & \ddot{\wth}_1 = 
    -\frac{2}{\wtr}\dwth_1 
    +\widetilde{\delta}_h\,, \; \ \; \ \qquad\quad\qquad\qquad
    \widetilde{\delta}_h = \dwts_1\dwth_0
    +\frac{2\wtw_0^2}{\widetilde{\mu}_0\wtr^2}\wth_0,
\nonumber  \\
    &\dwtm_1 = v\left(\wtr^2\dwth_0\dwth_1
    -\wtr^2\dwth_0^2\wts_1
    +\widetilde{\delta}_m\right)\,, \quad
    \widetilde{\delta}_m =  \frac{\wtw_0^2\wth_0^2}{\widetilde{\mu}_0}
    +\widetilde{\mu}_0\dot{\wtw}_0^2,
 \\
 & \dwts_1 = v\left(\frac{2\dot{\wtw}_0^2}{\wtr} +\frac{2\wtw_0^2\wth_0^2}{\widetilde{\mu}_0^2\wtr}\right),
 \nonumber  
\end{align}
which are subleading order terms of equation \eqref{eqn:EYMeqns} with respect to the expansion taken in equation \eqref{eqn:fnexpansions}.
These differential equations depend on the zeroth order quantities; we note that $\wts_0$, $\wth_0$ and $\wtm_0$ are directly calculated to be equation \eqref{eqn:hmzerothsol}, whereas $\tilde{w}_0$ is solved numerically, when $\Lambda$ and $h'_H$ are given, using the differential equation in equation \eqref{eqn:wEOMdetailed}. Hence, we know all zeroth-order quantities, and we can decide $\wth_1$, $\wtm_1$, $\wtw_1$, and $\wts_1$ from equation (\ref{eqn:EYMsubleading}). We first find $\dwth_1$ as
\begin{equation}
    \dwth_1(\wtr) = \frac{1}{\wtr^2}\int_1^{\wtr}d\wtrp\,\wtrpSsq\widetilde{\delta}_h(\wtrp).
\label{eqn:h1sol}
\end{equation}
Since $\sigma(r)$ goes to one when $\wtr\to\infty$, $\wts_1$ goes to zero as $\wtr\to\infty$, and,
\begin{equation}
    \wts_1(\wtr) = -\Delta \wts + \int_1^{\wtr} d\wtrp\, \dwts_1(\wtrp), \quad
    \Delta \wts \coloneqq \int_1^{\infty} d\wtr\, \dwts_1(\wtr).
\label{eqn:s1sol}
\end{equation}
Then, by taking the quantities $\dwth_1$ and $\wts_1$, given by equations \eqref{eqn:h1sol} and \eqref{eqn:s1sol}, we can solve $\wtm_1$ in equation \eqref{eqn:EYMsubleading} as
\begin{equation}
\scalemath{.95}{
    \frac{\wtm_1(\wtr)}{v}=\frac{1}{2v}+\int_1^{\wtr}d\wtrp\frac{h'_H}{\wtrpS}\int_1^{\wtrp}d\wtrpp\wtrppSsq\widetilde{\delta}_h(\wtrpp)
    +\int_1^{\wtr}d\wtrp\frac{h_H^{\prime 2}}{\wtrpS}\left(\Delta\wts-\int_1^{\wtrp}d\wtrpp\dwts_1(\wtrpp)\right)
    +\int_1^{\wtr}d\wtrp\widetilde{\delta}_m(\wtrp).}
\end{equation}
We are interested in $\wtm_1(\wtr\to\infty)$ because it corresponds to the mass of the black hole. The subleading contribution to the mass of the non-abelian black hole is expressed as
\begin{equation}
    \wtm_1(\wtr\to\infty)=\frac{1}{2}+\int_1^\infty d\wtr\,v \left(
    h'_H \wtr \widetilde{\delta}_h(\wtr) + h_H^{\prime 2}\left(1-\frac{1}{\wtr}\right)\dwts_1(\wtr)+\widetilde{\delta}_m(\wtr) \right).
\label{eqn:m1sol}
\end{equation}

Now that we have computed $\wth_1$, $\wtm_1$, and $\wts_1$, we can estimate the thermodynamic quantities of the black hole as
\begin{align}
    & Q \approx\, \frac{v}{G_N}\left(r_H^2 h'_H + \int_1^{\infty}d\wtr\ \wtr^2 \widetilde{\delta}_h(\wtr) \right)\frac{\tau_3}{2} ,\nonumber \\
    & T \approx\, \left.\frac{r_H}{4\pi G_N}\,\dot{\widetilde{\mu}}_0\right|_{\wtr=1}+\left.\frac{1}{4\pi G_N r_H}\left(\dot{\widetilde{\mu}}_1-\dot{\widetilde{\mu}}_0\Delta\wts\right)\right|_{\wtr=1},
    \nonumber \\
    & M \approx\, \frac{-\Lambda+3vh_H^{\prime\; 2}}{6G_N}\,r_H^3 + \frac{\wtm_1(\wtr\rightarrow\infty)}{G_N}\,r_H ,
\label{eqn:NAMQTsublead}
\end{align}
where $\wtr =r/r_H$ and the mass $M$ is evaluated in the infinite radius limit, provided from the value of $m(r)$ at $r\to\infty$.
Finally, the Helmholtz free energy of black hole with non-abelian hair is given by,
\begin{align}
\begin{aligned}
    F&=M-TS\\
    &=r_H^3\widetilde{F}_0 + \frac{r_H}{G_N}\left(\frac{1}{4}+ v\int_1^{\infty} d\wtr\ \Delta\wtm(\tilde{r}) + \frac{1}{4}(-\Lambda + vh_H^{\prime\; 2})\Delta\wts \right)\,,
\label{hairlyfreeenergy}
\end{aligned}
\end{align}
where $v\Delta \wtm$ is the integrand of the equation \eqref{eqn:m1sol} and
\begin{equation}
        \widetilde{F}_0=\frac{1}{G_N}\left(\frac{1}{12}\Lambda + \frac{3}{4}vh_H^{\prime\; 2}\right) \,.
\end{equation}

\begin{figure}[H]
	\centering
	\includegraphics[width=0.8\textwidth]{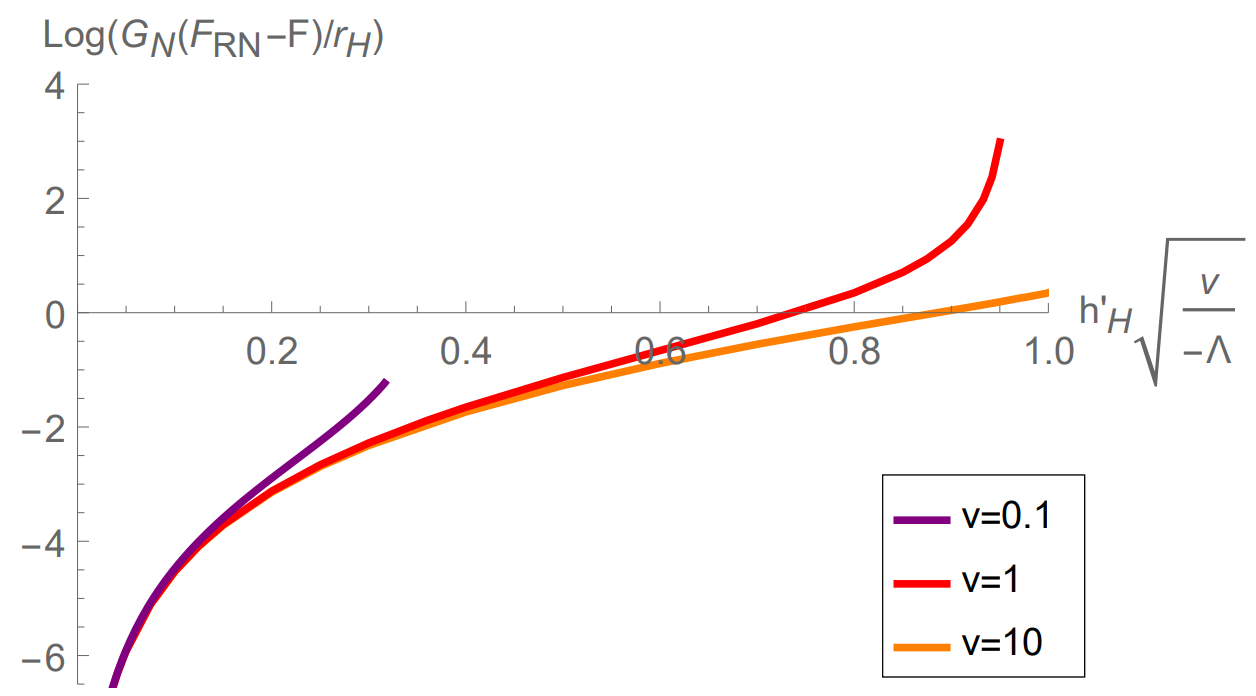}
	\caption{$\log\left[ G_N(F_{RN}-F)/r_H\right]$ as a function of ${h'_H\sqrt{v/-\Lambda}}$ when $\Lambda=-1$}
	\label{fig:deltam}
\end{figure}

Let us compare this with the free energy of the $U(1)$ RN black hole. 
For the solution to have the same temperature and charge, 
the radius of the horizon of the RN black hole must be
\begin{equation}
    r_{H,RN}=r_H+\frac{\Delta\wtr}{r_H}\,, \quad \Delta\wtr =\frac{-(-\Lambda+vh_H^{\prime\; 2})\Delta\wts +2vh'_H\int_1^{\infty}d\wtr\ \wtr^2 \widetilde{\delta}_h(\wtr)}{-\Lambda+3vh_H^{\prime\; 2}}\,.
\end{equation}
The free energy is then,
\begin{equation}
    F_{RN}= r_H^3 \widetilde{F}_0 + \frac{r_H}{G_N} \left(
    \frac{1}{4} + vh'_H\int_1^\infty d\wtr\ \wtr^2 \widetilde{\delta}_h(\wtr)
    + \frac{1}{4}(-\Lambda + vh_H^{\prime\; 2})\Delta\tilde{\sigma}
    \right)\,.
    \label{abelianfreeenegy}
\end{equation}
We remark again that the two free energies in equations \eqref{hairlyfreeenergy} and \eqref{abelianfreeenegy} have same leading behavior. By taking the difference of the two free energies, we obtain
\begin{equation}
    F_{RN} - F = \frac{v\,r_H}{G_N}\left( h'_H\int_1^\infty d\wtr\ \wtr^2\widetilde{\delta}_h(\wtr) -\int_1^{\infty}d\wtr\ \Delta\widetilde{m}(\wtr)\right)\, , 
    \label{eqn:Fdifference}
\end{equation}
which comes from the $1/T$ correction.
When we numerically calculate this difference, it has strictly positive value as shown in Figure \ref{fig:deltam}.
Therefore, the black hole with non-abelian hair has a smaller free energy
and is thermodynamically preferred over the $U(1)$ RN black hole at finite temperature. 

\subsection*{Acknowledgements}
We thank J. Bhattacharya, D. Harlow, G. Horowitz, T. Melia, S. Minwalla, S. Pal, D. Simmons-Dufffin, Z. Sun, T. Takayanagi, and Z. Zhang for discussion.
This work is supported in part by the US Department of Energy
under the award number DE-SC0011632. 
M.J.K.~is supported in part by the Sherman Fairchild Postdoctoral Fellowship.
H.O.~is supported in part by the World Premier International Research Center Initiative,
MEXT, Japan, and by JSPS Grant-in-Aid for Scientific Research 20K03965. 
This work was performed in part at the Aspen Center for Physics, which is supported by NSF grant PHY-1607611.

\bibliography{references}{}
\bibliographystyle{nosortnodoi}
\end{document}